\documentclass[epj,final]{svjour}% referee final
\usepackage{graphics}
\begin{document}
\title{The hadronic potential at short distance}
\author{Hans-Christian Pauli}
%
%  Comments: LaTeX2e, 4 pages, 3 figures, 0 tables, 10 references. 
%
\institute{Max-Planck-Institut f\"ur Kernphysik, D-69029 Heidelberg }
\date{15 December 2003}  
\abstract{%
    A fictitious discussion is taken as a point
    of origin to present novel physical insight into the nature of 
    gauge theory and the potential energy of QCD and QED 
    at short distance. Emphasized is the considerable freedom
    in the cut-off function which eventually can modify the Coulomb 
    potential of two charges at sufficiently small distances. 
    Emphasized is also that the parameters of the
    regularization function (the ``cut-off scale'')
    should not be driven to infinity but kept constant
    in line with the modern interpretation of renormalization theory.
    The paper restricts to general aspects.
    The technical paraphernalia and the comparison with experiment are
    shifted to a sequence of 4 subsequent stand-alone and 
    sufficiently small papers to be published immediatelely hereafter.  
} 
\maketitle
\section{Introduction: Highlights of a discussion}

In the aftermath of recent conference contributions \cite{Pau03a},
I had to survive serious discussions whose highlights 
should be interesting for a larger audience. 
In a way, the arguments of a fictitious anonymous represent
the state of the art public opinion of the high 
energy physics community. 
I let play him the role of the devils advocate.
\noindent
Here are some of his points: 
\begin{enumerate} 
\item  
       QCD interactions at short distances are
       scale invariant~-- up to the logarithmic asymptotic
       freedom corrections predicted from the running coupling.
       This near conformal scaling of QCD is
       well tested in quark-quark scattering and
       other jet physics measurements at colliders.
\item  \label{Brodsky:scale}
       Certainly $\Lambda_{QCD}$ sets the scale
       of the hadronic parameters.
       It is relevant to QCD interactions at large distances.
\item 
       Evidence is accumulating that 
       the running coupling~-- as defined from the  
       Landau gauge gluon propagator or observables~-- 
       is well regulated at small $Q^2$.
       Some relevant papers are 
       \cite{vonSmekal:1997is,Howe:2002rb,Mattingly:1994,Brodsky:2002nb}.
\item  \label{Brodsky:Fourier}
       It might be worthwhile to look at the Fourier transform of 
       ${\alpha_s(Q^2)}/{Q^2}$ where 
       $\alpha_s(Q^2)$ is the Shirkov form,   
       as given for example in the paper \cite{Baldicchi:2002}.
\item  \label{Brodsky:Yukawa}
       Pauli uses a cut-off $\lambda$ to motivate a
       modification of the short distance $r \to 0$ QCD interaction.~---
       Let us compare the Yukawa ${1}/{(Q^2+\lambda^2)}$ 
       and Coulomb ${1}/{Q^2}$ interactions. 
       They are identical at large $Q^2$.~---  
       Their respective Fourier transforms potentials are 
       the ${\mathrm{e}^{-\lambda r}}/{r}$ and $1/r$.   
       They are identical at short distances $r \to 0$!
\end{enumerate}

\noindent 
Here is a summary of my points: 
\begin{enumerate} 
\item 
       Relating large momentum transfers to short distances 
       is often only a figure of speech.
       The relation between a scattering amplitude $T(Q^2)$ and
       an interaction $V(r)$ is a highly non-trivial matter.
\item 
       Colliders measure the scattering amplitude $\alpha_s(Q^2)/Q^2$ 
       at large $Q^2$.
       The interaction is the Fourier transform
       of ${\alpha_s(Q^2)}/{Q^2}$,   
       all $Q^2$ are needed, and not only the large ones. 
\item 
       We should be more careful about our terms of speaking,
       and specify by operational prescriptions what we mean by 
       `interaction' and/or `potential'.
\item 
       The $q\bar q$ potential at short distance 
       simply cannot go like $\alpha_s/r$, whatever the numerical value
       of the strong coupling `constant' $\alpha_s$ is. 
\item 
       The literature has too many $\Lambda$'s!
       The same mathematical symbol refers to completely different physics.
\end{enumerate}

\noindent
In this short note, I will work out to some detail why and to which extent 
I will arrive at conclusions different from the public opinion.

I use this opportunity as the port of entry for presenting novel insights 
on the nature of the gauge field interaction at short distance. 
In \cite{Pau03b} 
I will present a possible solution to the problem of the non 
perturbative renormalization in a gauge theory. 
In \cite{Pau03c}, the technical details of the fine and hyperfine 
interaction on the light cone will be discussed. 
In \cite{Pau03d} and \cite{Pau03e}
I will calculate the ground state masses and mass spectra
of all flavor off diagonal pseudo scalar and pseudo vector mesons, 
analytically, for a linear and for a quadratic potential.

\section{Taking Fourier transforms}
\label{sec:2}
Collider physics is consistent with asymptotic freedom,  
\begin{eqnarray}
   \alpha_s(Q^2)= \frac{4\pi}{\beta_0\mathrm{ln}( Q^2/\Lambda^2_{QCD} )}  
\,,\label{eq:1}\end{eqnarray}
at sufficiently large $Q^2$, with the famous $\beta_0=11n_c-2n_f$. 
What is its Fourier transform?

Let us be specific. Consider  
\begin{eqnarray}
   V(r) &=& -\frac{1}{2\pi^2}\int\!d^3 q\;
   \frac{\alpha_c(q^2)}{q^2}\;
   \mathrm{e}^{-i\vec q \vec r}
,\label{eq:2}\end{eqnarray}
where $\alpha_c(q^2)=\frac43 \alpha_s(q^2)$ given by Eq.(\ref{eq:1})
with the Feynman four momentum transfer $Q^2$ substituted by 
the three momentum transfer $q^2=(\vec k - \vec k')^2$.
Integrating over the angles gives
\begin{eqnarray}
   V(r) &=& -\frac{1}{\pi}\int_{0}^{\infty}\!d q\;q^2\;
   \frac{\alpha_c(q^2)}{q^2}\;
   \int_{-1}^{+1}\!\!\!\!d (\cos\theta)\;
   \mathrm{e}^{-iq r \cos\theta}
\nonumber\\ &=& -
   \frac{2}{\pi}\int_{0}^{\infty}\!d q\;{\alpha_c(q^2)}\;
   \frac{\sin qr}{qr}
.\label{eq:3}\end{eqnarray}
But this integral does not exist!

The problem is as absent for short distances $r\to0$, as long
as $\alpha_c(q^2)$ is integrable. The available `effective'
$\alpha_c(q^2)$ in 
\cite{vonSmekal:1997is,Howe:2002rb,Mattingly:1994,Brodsky:2002nb} 
may or may not allow for $\int_{0}\!d q\;{\alpha_c(q^2)}$.

The problem resides at the upper limit.
A Fourier transform of a function is defined only
if its \textit{limes superior} exists, that is,
it exists only if
\begin{eqnarray*}
   \int^{\infty}\!\frac{d q}{q}\;{\alpha_c(q^2)}\;
\end{eqnarray*}
is a well defined expression. But Eq.(\ref{eq:3}) it not well defined.
It diverges at the upper limit, since:
\begin{eqnarray}
   \int\!\frac{d q}{q \ln q} = \ln \ln q
\,.\end{eqnarray}
Knowing that asymptotic freedom is too weak a regulator,
I have not even attempted in the past, to carry out 
point~\ref{Brodsky:Fourier} of the advocate. 
Baldicchi and Prosperi  \cite{Baldicchi:2002} 
also know about the problem, but they veil it by 
inserting a linear potential as a regulator,  by hand
and out of desperation.

One concludes that the integral, Eq.(\ref{eq:2}), 
\emph{must be regulated}. What are possible alternatives?

Cutting-off at the upper limit \`a la $\int^{q_0}\!d q$ 
makes not much sense sense. 
A sharp cut-off with the step function $R(q)=\Theta(q^2-\lambda^2)$ 
generates uncontrollable oscillations. 
A soft regulator function, as for example
\begin{eqnarray}
   R(q) = \frac{\lambda^2}{\lambda^2+q^2} 
\,,\label{eq:5}\end{eqnarray}
satisfies however all important requirements, among them
$R(q)\Longrightarrow 1$ for $\lambda\Longrightarrow\infty$
and no extra physical dimensions. It is dimensionless.

In consequence one replaces Eq.(\ref{eq:2}) by
\begin{eqnarray}
   V(r) &=& -\frac{1}{2\pi^2}\int\!d^3 q\;
   \frac{\alpha_c(q^2)}{q^2}\;R(q)\; 
   \mathrm{e}^{-i\vec q \vec r}
,\end{eqnarray}
and Eq.(\ref{eq:3}) by
\begin{eqnarray}
   V(r) &=& -
   \frac{2}{\pi}\int_{0}^{\infty}\!d q\;{\alpha_c(q^2)}\;R(q)\;
   \frac{\sin qr}{qr} 
.\end{eqnarray}
One faces now a well defined mathematical problem.

But one faces another problem: 
The integrand behaves like:
\begin{eqnarray*}
   \frac{4\pi}{\beta_0\mathrm{ln}(q^2/\Lambda^2_{QCD})}\;
   \frac{\lambda^2}{\lambda^2+q^2}\;
   \frac{\sin qr}{qr} 
.\end{eqnarray*}
The large q behavior is dominated completely by the regulator 
and the rapid oscillations of the sine function. 
As compared to them, the logarithm is hyper slowly varying. 
A mathematician would treat the problem by replacing
the slowly varying term by a constant $\alpha'_c$ 
and take it out of the integral. 
(He would call this step the saddle point approximation.) 
He thus would get
\begin{eqnarray}
   V(r) &=& -
   \frac{2\alpha'_c}{\pi}\int_{0}^{\infty}\!d q\;R(q)\;
   \frac{\sin qr}{qr} 
\\ &=& \displaystyle
   \alpha'_c\Big[\frac{\mathrm{e}^{-\lambda r}}{r} - 
   \frac{1}{r}\Big] 
\,.\label{eq:9}\end{eqnarray}
Here are the Coulomb's and Yukawas which the advocate 
refers to in his point~\ref{Brodsky:Yukawa}.
He is right, of course, in stating that 
they are identical at short distances $r \to 0$.

The notion `identical' is misleading:
The leading singularities cancel, but still $V(r\to 0)$ is non-zero.
The systematic expansion for short distances rather gives  
\begin{eqnarray}
   V(r) &=& -a + b\;r
\,,\label{eq:10}\end{eqnarray}
plus possible higher terms to which I shall come below.

The numerical values of $a$ and $b$ are less important for the argument.
But for completeness, here they are:
\begin{eqnarray}
   a &=& \alpha'_c\;\lambda 
\,,\\
   b &=& \alpha'_c\;\lambda^2/2  
\,.\end{eqnarray}
The considerations, which lead to Eq.(\ref{eq:10}), also 
highlight another aspect. 
The only piece of non ambiguous experimental information 
is the coupling constant at very high momentum transfer:
The `running' constant at the $Z$ mass $M_Z\simeq 91.2\mbox{ GeV}$.
Its experimental value is $\alpha_s(M_Z)\simeq 0.118$.
The coupling constant $\alpha_s'=\alpha_s(q_0)$ 
can thus be taken at comparatively large values of $q_0$.
The argument of $\alpha'_s=\alpha_s(Q=0)$ in \cite{Pau03a} 
is false, since $\alpha'_s=\alpha_s(q_0)$, but the preliminary value 
\begin{eqnarray}
   \alpha'_s = 0.1695 
\,,\end{eqnarray}
taken from \cite{Pau03a},
is a useful first guess subject to later re adjustment. 
The numerical value of $\alpha'_s$ is comparatively small, 
much smaller in any case than the often quoted value $\alpha'_s = 0.7$.
Setting the scale, like in \cite{Pau03a},
\begin{eqnarray}
   \lambda = 200\mbox{ MeV} 
\,,\end{eqnarray}
determines all parameters.

The above argument shows that the asymptotic scale $\Lambda_{QCD}$ 
in Eq.(\ref{eq:1})
is conceptually different from the regularization scale $\lambda$
in Eq.(\ref{eq:5}).
It is $\lambda$, not $\Lambda_{QCD}$, which sets the scale of 
the problem~-- contrary to point~\ref{Brodsky:scale} of the advocate.
The confusion arises since both scales are often quoted with 
the same numerical value of about $200\mbox{ MeV}$,  
corresponding to a length scale of about $1\mbox{ fm}$.

%
%%%%%%%%%%%%%%%%%%%%%%%%%%%%%%%%%%%%%%%%%%%%%%%%%%%%%%%%%%%%%% beg figure
\begin{figure}[t] 
   \resizebox{0.48\textwidth}{!}{\includegraphics{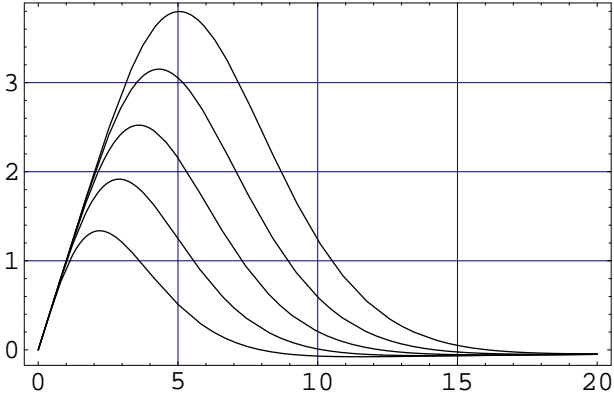}
}\caption{ 
   The dimensionless Coulomb potential $W_N(y;0,1,0)$ is plotted versus 
   the radius parameter $y=\lambda r$. The different $N=4,5,6,7,8$ 
   are drawn by the curves from bottom to top. 
}\label{fig:lin}\end{figure}
%%%%%%%%%%%%%%%%%%%%%%%%%%%%%%%%%%%%%%%%%%%%%%%%%%%%%%%%%%%%%% end figure
%
%
\section{The generalized regulator function}

It should be emphasized that the limit $\lambda\to\infty$
may not be taken in Eq.(\ref{eq:9}): For $\lambda\to\infty$, 
one is back to the undefined problem of Eq.(\ref{eq:3}).

The same conclusion will also be reached with the 
renormalization group equations \cite{Pau03b}.
The renormalization group equations \cite{Pau03b} 
require also that the regulator function has well
defined derivatives with respect to $\lambda$.
This excludes the step function of the sharp cut-off 
from the class of admitted regulator functions. 
The theta function is a distribution with only ill defined derivatives.
Having understood these essentials, one can phrase things  
in a very simple way.

Do the above statements imply that the `soft' 
regulator in Eq.(\ref{eq:5}) is the only one admitted?~--- 
Of course not. 

In fact, one has a large class 
of `generalized regulator functions' \cite{Pau03a}: 
\begin{eqnarray}
   R(q) = \left[1+\sum_{n=1}^{N}
   (-1)^n s_n \lambda^n\frac{\partial^n}{\partial\lambda^n}\right]
   \frac{\lambda^2}{\lambda^2+q^2} 
\,.\label{eq:35}\end{eqnarray}
The partials $\lambda^n\,\partial^n/\partial\lambda^n$ are 
dimensionless and independent of a change in $\lambda$.
The arbitrarily many coefficients $s_1,\dots,s_N$ are 
renormalization group invariants and, as such, subject to be
determined by experiment.

The generalized regulator yields immediately the generalized Coulomb potential
\begin{eqnarray}
   V(r) &=& -\frac{\alpha_c}{r} \Big[1+\sum_{n=1}^{N}
   (-1)^n s_n \lambda^n\frac{\partial^n}{\partial\lambda^n}\Big]
   \Big(1-\mathrm{e}^{-\lambda r}\Big) 
\nonumber\\ &=& \phantom{-}
   \frac{\alpha_c}{r}\Big[
   -1+ \mathrm{e}^{-{\lambda r}}\sum_{n=0}^{N}s_n({r\lambda})^n \Big]   ,
\label{eq:36}\end{eqnarray}
with $s_0\equiv1$.
This result illustrates an other important point:
The power series in front of the exponential are nothing 
but a spelled out version of Laguerre polynomials. 
Laguerre polynomials are a complete set of functions.
Eq.(\ref{eq:36}) is thus potentially
able to reproduce an arbitrary function of $r$. 
The description in terms of a generating function, 
as in Eqs.(\ref{eq:35}) or (\ref{eq:36}), 
is therefore \emph{complete}.

%
%%%%%%%%%%%%%%%%%%%%%%%%%%%%%%%%%%%%%%%%%%%%%%%%%%%%%%%%%%%%%% beg figure
\begin{figure}[t] 
   \resizebox{0.48\textwidth}{!}{\includegraphics{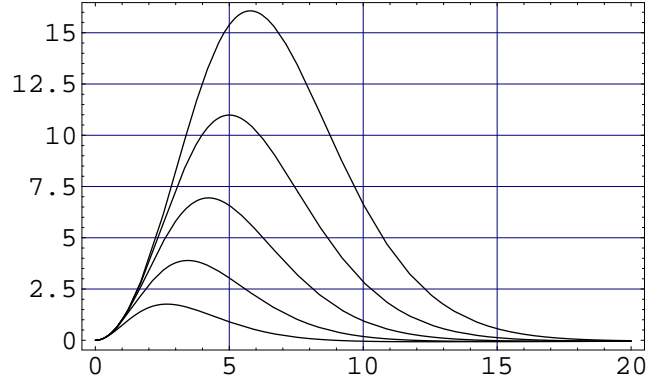}
}\caption{ 
   The dimensionless Coulomb potential $W_N(y;0,0,1)$ is plotted versus 
   the radius parameter $y=\lambda r$. The different $N=4,5,6,7,8$ 
   are drawn by the curves from bottom to top. 
}\label{fig:osc}\end{figure}
%%%%%%%%%%%%%%%%%%%%%%%%%%%%%%%%%%%%%%%%%%%%%%%%%%%%%%%%%%%%%% end figure
%
\textbf{The large number of parameter} in Eq.(\ref{eq:36}) 
can be controlled by the following construction:
The coefficients $s_n$ in Eq.(\ref{eq:36})
are expressed in terms of only three parameters $a$, $b$, and $c$, by 
\begin{eqnarray}
    s_n = 
    \frac{1}{n!} + \frac{a}{(n-1)!} + \frac{b}{(n-2)!} + \frac{c}{(n-3)!} 
\,.\label{eq:37}\end{eqnarray}
The first few coefficients are then explicitly
\begin{eqnarray}
   \begin{array}{rc rc rc rc rc rc }
    s_0 &=& 1    &,& \\ 
    s_1 &=& 1 &+&  a &,& \\ 
    s_2 &=& \frac{1}{2}    &+& a &+& b &,& \\  
    s_3 &=& \frac{1}{6}    &+& \frac{a}{2}    &+& b            &+&  c &,& \\ 
    s_4 &=& \frac{1}{24}   &+& \frac{a}{6}    &+& \frac{b}{2}  &+&  c &.&   
   \end{array} 
\end{eqnarray}
As a consequence, the dimensionless Coulomb potential 
depends on $r$ only through the dimensionless combination $y=\lambda r$:
\begin{eqnarray}
   W_N(y;a,b,c) = 
   \frac{V(r)}{\alpha_c\lambda} =
   \frac{1}{y}\Big(-1+ \mathrm{e}^{-y}\sum_{n=0}^{N}s_ny^n\Big) 
\,.\label{eq:39}\end{eqnarray}
In the near zone, it is a quadratic function of $y$, 
\begin{eqnarray}
   W_N(y;a,b,c) = a + by + cy^2
\,,\label{eq:40}\end{eqnarray}
independent of the value of $N$. 
The remainder starts at most with power $y^{N+1}$.
A value of $a=c=0$ and $b=1$ should 
yield a linear set of functions $W_N(y;0,1,0) \simeq y$
in the near zone. As shown in Fig.~\ref{fig:lin} 
with the original Mathematica plot,
this happens to be true for surprisingly large values of $y$,
\textit{i.e.} not only for $y\ll 1$. 
The value of $N$ essentially controls the height of a barrier.
Similarly, $W_N(y;0,0,1)=y^2$ generates a set of functions
which are strictly quadratic in the near zone. 
Again, $N$ controls the height of a barrier.

\begin{figure}
\resizebox{0.48\textwidth}{!}{\includegraphics{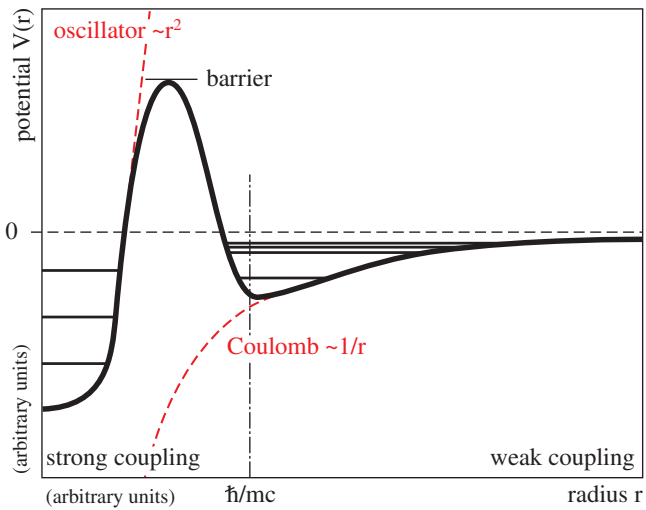}
}\caption{ 
   Schematic behavior of the renormalized Coulomb potential,
   see also the discussion in the text.
}\label{fig:V(r)schem}\end{figure}

\textbf{The physical picture} which develops is illustrated 
in Fig.~\ref{fig:V(r)schem}.
In the far zone, for sufficiently large $r$,
the potential energy coincides with the conventional Coulomb potential
$-\frac{\alpha_c}{r}$. 
Since the potential is attractive, it can host bound states
which are probably those realized in weak binding.
In the near zone, for sufficiently small $r$,
the potential behaves like a \emph{power series} $c_0+c_1r+c_2r^2$
which potentially can host the bound states of strong coupling,
provided the actual parameter values allow for that.
In the intermediate zone, the actual potential must interpolate
between these two extremes, since Eq.(\ref{eq:36}) is an analytic
function of $r$. Most likely this is done by developing 
a \emph{barrier of finite height}, 
depending on the actual parameter values.
The onset of the near and intermediate regimes must occur
for relative distances of the quarks, which are comparable
to the Compton wave length associated with their reduced mass.
If the distance is smaller, one expects deviations from the classical 
regime by elementary considerations on quantum mechanics, indeed.

\section{Conclusions and remaining mysteries}
Once the arguments in Sec.~\ref{sec:2} are accepted 
for Quantum Chromodynamics (QCD), one must accept
them also for Quantum Electrodynamics (QED).
The integral in Eq.(\ref{eq:3}) for QED diverges even stronger
than for QCD. I thus arrive at the incredible conclusion 
that the potential energy between a muon and an electron
is linear (or quadratic!) for sufficiently short distance.

Do I want to overthrow everything?~--- As a matter of fact: no!
The scales are such that the hadronic length scale 
$R_H=\hbar/\lambda c\sim 1\mbox{ fm}$ 
is very much smaller than the Bohr scale 
$R_B=\hbar/m_ec\cdot 1/\alpha\sim 386\cdot 137\mbox{ fm}$.
The electron simply `does not see' the details at short distance
except for a small perturbation. 
We know that from the hydrogen atom where the finite size of the proton
gives only a small finite size correction.

The second important conclusion is that linear (or quadratic!)
confinement can not go on forever. 
Confinement must be short distance phenomenon, 
in sharp contrast to the teleological or theological 
beliefs of the community.
The hadronic potential allows for continuous spectra.

The only mystery remaining is that Lattice Gauge Theories 
and their representatives
insist on a singularity at short distance,
\begin{eqnarray}
   V(r) &=& -\frac{\alpha'_c}{r} + b\;r \neq -a + b\;r 
\,,\label{eq:12}\end{eqnarray}
although inherently to the method, Lattice Gauge Theory
can not carry out calculations \emph{at the singularity} $r=0$.
The value of the constant $a$ can take arbitrarily large values,
but it must be finite, by definition. 
I have insufficient working knowledge to comment 
any further on these questions.

``Subtle is the Lord, but nasty is he not.''
%
%%%%%%%%%%%%%%%%%%%%%%%%%%%%%%%%%%%%%%%%%%%%%%%%%%%%%%%%%%%%%% beg figure
%%%%%%%%%%%%%%%%%%%%%%%%%%%%%%%%%%%%%%%%%%%%%%%%%%%%%%%%%%%%%% end figure
%

\acknowledgement
\textbf{Acknowledgement.}
I thank my good friend and light cone mentor Stan Brodsky
for an intense discussion by e-mail which popped up 
in early October 2003 and went over 10 rounds. 
I thank as well my friend Bogdan Povh for his continuous support
over the years, and the encouragement 
to switch from nuclear physics to the light cone.
I beg pardon to quote myself so often. 
I hate it that many authors refer to all their own papers 
ever written. But here, I have to do it by technical reasons.
Sometimes it takes courage to be provincial, as Bogdan would say.

\end{document}